# Accepted Paper (No. 106)

**Paper Title:** Segregated FLS Processing Cores for V/STOL Autonomous Landing Guidance Assistant System Using FPGA

**Conference Name:** ICNS Horizons 2021 Agility and Sustainability: Embracing Rapid Evolution and Uncertainty

**Paper Author:** Hossam O. Ahmed ( https://www.linkedin.com/in/hossamomar/ )

**IEEE ICNS 2021 Sponsors**

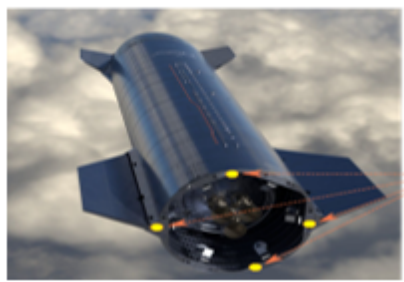

The four HOA units

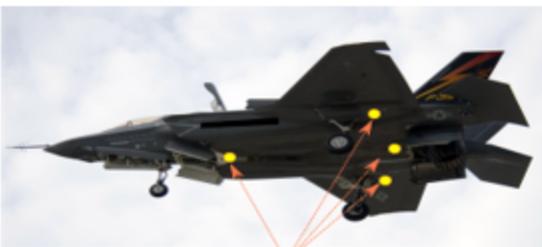

The four HOA units

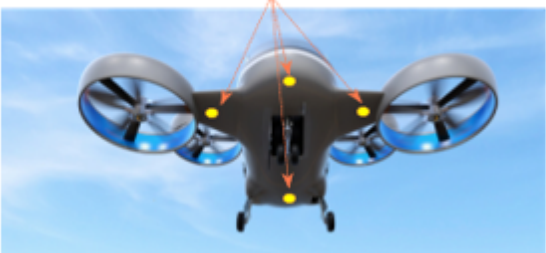

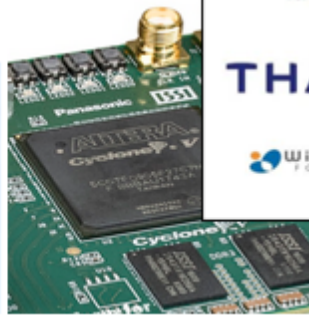

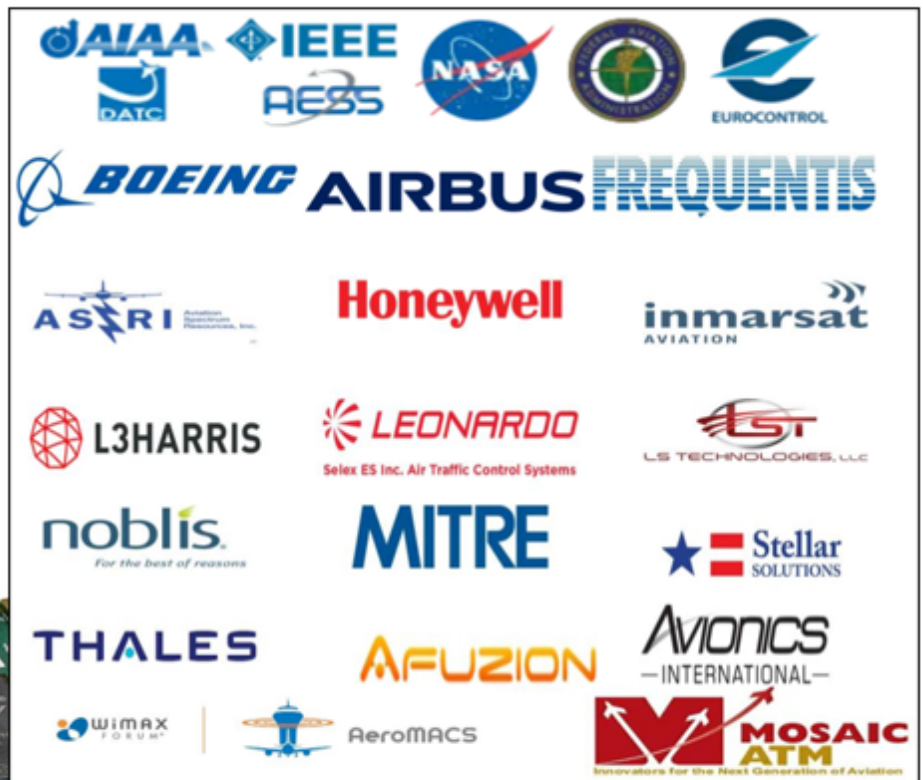

# Segregated FLS Processing Cores for V/STOL Autonomous Landing Guidance Assistant System using FPGA.

*Hossam O. Ahmed, College of Engineering and Technology, American University of the Middle East, Kuwait.*

## Abstract

It is highly predicted that the roads and parking areas will be extremely congested with vehicles to the point that searching for a novel solution will not be an optional choice for conserving the sustainability rate of the overall humanity's development growth. Such issue could be overcome by developing modified generations of the Urban Air Mobility (UAM) vehicles that essentially depend on the Vertical and/or Short Take-Off and Landing (V/STOL) feature to increase the efficiency of landing capabilities on limited- space parking areas. The complexity of integrating an efficient and safe V/STOL feature in such UAM vehicles is notably difficult comparing with the conventional and normal techniques for landing and take-off. The efficient V/STOL feature should be carried out by a complete and collaborative Cyber-Physical System (CPS) processing architecture, such as the CPS-5C architecture. In this paper, we only proposed two CPS-5C physical layers of a V/STOL Autonomous Landing Guidance Assistant System (ALGAS2) processing unit to increase the reliability of the vertical landing mechanism.

The proposed V/STOL – ALGAS2 system depends on Fuzzy Logic System (FLS) as the advanced control unit. Furthermore, the proposed ALGAS2 system depends on four symmetric and segregated processing ALGAS2 cores that processing the data in a fully parallel and independent manner to enhance many essential security and safety factors for the futuristic UAM vehicles. The proposed ALGAS2 digital circuits architecture has been designed using MATLAB and VHDL. Also, it has been further analyzed for the implementation and validation tests using the Intel Altera OpenVINO FPGA board. The proposed ALGAS processing unit attained a maximum computational processing performance of about 21.22 Giga Operations per Seconds (GOPS).

## Introduction

The tremendous increase in the world population makes it almost impossible to rely on the existing transportation system's solutions. Many attempts and proposals of innovative ideas to allocate well-defined multi-level air pathways have been introduced. Subsequently, there is a must for escalating the precautions of air safety standards from many different aspects [1-4]. One of the key aspects, that requires to be focused on, is the autonomous landing safety factor of such Urban Air Mobility (UAM) vehicles [5]. Moreover, UAM vehicles will depend on Vertical and/or Short Take-Off and Landing (V/STOL) mechanism [6].

The V/STOL mechanism needs to be extremely efficient against many expected challenges such as the inexpert pilots and the limited and/or unpredicted landing spaces such as in emergency cases. Enhancing the efficiency of the autonomous V/STOL mechanism is also beneficial in the equivalent space and military applications [7]. In this paper, we proposed a generic four symmetric and segregated Autonomous Landing Guidance Assistant System (ALGAS) cores support the autonomous landing ability of any V/STOL- based air vehicles. This proposed system could also mitigate the landing failures due to either the human intervention errors or these which comes from the consequences of communication links issues. Such promising feature could be achieved by the aggregation the vital sensory information from four different Of Hybrid Obstacle Avoidance (HOA) units, which are located on the bottom side of any V/STOL-based air vehicles such as in Figure1. Subsequently, all these sensory data transferred to proper ALGAS processing cores. In section II, we review the related works and similar research and contributions. In section III, we gave an overview of the proposed ALGAS2 digital circuits architectures. In section IV, we discussed the experimental results obtained after the synthesis process of the proposed ALGAS2 digital circuits architecture using Intel Quartus Prime. In section V, we presented the conclusion and the future works.

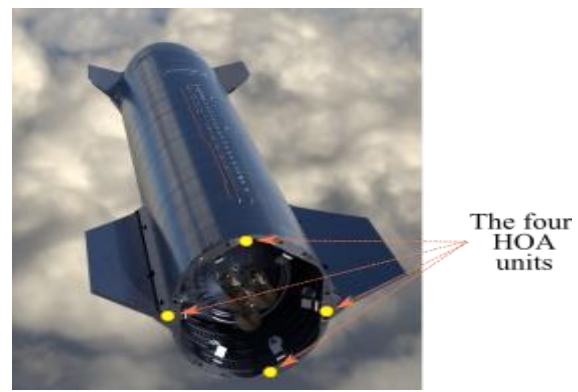

**Figure 1. A hypothetical integration of quad Hybrid Obstacle Avoidance (HOA) modules on the Masterpiece SpaceX Starship rocket.**

## Related Work

The pivot point, that will guarantee the success of applying UAM worldwide, is directly connected to the well understanding and the degree of efficiency for applying safety factors. Novel and adaptable safety standards should be integrated with the existing safety regulations for Single European Sky ATM Research (SESAR) and the Next Generation (NextGen) air transportation systems [8, 9]. Different research studies explained the significant rule of applying advanced control systems and Artificial Intelligence (AI) algorithms to enhance variety of capabilities of UAM systems [10-12]. This includes propositions related to trajectory conflict detection and avoidance mechanisms, smart path planning, and configurable air traffic techniques [13-15]. However, one of the most critical safety concerns for the next UAM systems will be related to the efficiency of the autonomous landing process.

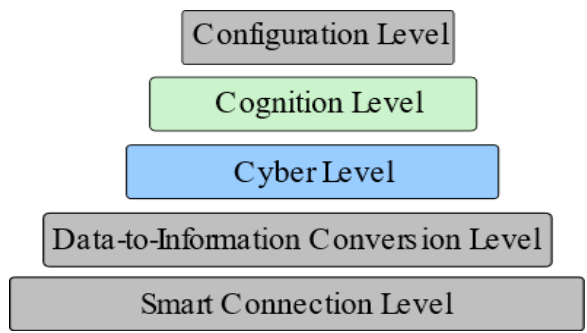

**Figure 2. A graphical representation of the CPS-5C hierarchy.**

This landing safety concern arises due to the expected wide gap and shortage between the tremendous number of the predicted UAM vehicles in the market and the availability of expert pilots that we can be afforded. Also, we have the problem of unpredicted landing terrains especially under emergency landing circumstances [7, 16].

One of the prominences AI algorithms that has been proposed to overcome many UAM issues, such as autopilot dynamics, visual human tracking drone systems, is the Fuzzy Logic System (FLS) [17, 18]. The FLS is a very robust advanced control system that could handle any uncertainties and noises in electronic systems [19, 20]. It has been proven in [21, 22] that FLS can optimize the safety concerns in self-landing and hovering missions for UAM vehicles. Also, it could be combined with neural networks to enhance the altitude control as proposed in [23].

An interesting multicore, systolic, and real-time processing architecture using Field Programable Gate Array (FPGA) chip has been proposed in [7] to increase the efficiency of the autonomous self-landing process for future UAM vehicles. Moreover, this goal achieved by adopting parallel and complex sensory fusion processing architecture. This system is interactively dealing with Multiple Sensor Nodes (MSN) that are both frequency spectrum and spatially separated on the bottom side of an UAM. This proposed idea in [7] surpassed the conventional single sensor-array based-solutions for UAM landing process in terms of improving the accuracy and safety concerns only in the area of self-landing automation ability.

It is so essential to have a design that can efficiency comply the specific assigned tasks. However, the ability of the subsystem to be easily integrated with high-level autonomous systems is a key factor for sustaining the scalability and agility in such projects. One of the best system architectures for processing and controlling the autonomous systems is the Cyber Physical System (CPS-5C). The prominence of the CPS-5C architecture is the high-degree of flexibility to work in different autonomous applications such as in medical applications, automotive, smart cities, industrial, consumer, energy, robotics, infrastructure, transportation, and military [24-26]. As demonstrated in Figure 2, the CPS-5C architecture consists of 5 layers. The first layer, the smart connection level, concerns about the physical interconnections of the proper sensory modules and the communications transceiver. The second layer is responsible for extracting the vital data from the given sensor information by using control algorithm and degradation and performance measurements. The Cyber level layer is responsible for clustering and variation detection. All the tasks which related to human monitoring and interfacing, or collaborative diagnostics and decision-making is under the jurisdiction of the cognition level. The last layer is responsible for the system configuration and self-optimization for disturbance or any failure to the system [27, 28].

## The proposed ALGAS Processing Unit

This paper is focusing on the development of a novel ALGAS2 system that could promote the attempts of finding suitable and optimum safety standards for the next generations of Urban Air Mobility (UAM) vehicles. The proposed safety paradigm could be categorized into three different sectors. The 1st safety sector is related to guarantee the scalability of the proposed ALGAS2 unit. This could be measured by its ability to be integrated and collaborating with other subsystems in the top-level system hierarchy such as the CPS-5C architecture. The main functionality of the mentioned CPS-5C architecture, in this paper, is for enhancing safety factor only during the V/STOL operations. Also, the proposed ALGAS2 system in this paper is only covering two of the five CPS-5C architecture, which are the cyber level and the cognition level.

The proposed ALGAS2 system is consists of four ALGAS2 cores as depicted in Figure 3. These four ALGAS2 cores are responsible of forming a complete autonomous V/STOL assistant system at the congestion level. The CPS-5C cyber level, in the proposed ALGAS2 system, is represented by the formation between the Active Hub (AH) unit and the four Network Interface (NI) units. The main functionality of this layer is organizing the flow of data between the ALGAS2 cores, and to prevent and collision of information to be shared among them. Also, the digital circuits architecture of the proposed cyber and cognition layers can be easily integrated with the other 3 CPS-5C layers in the future progression of the project.

The 2nd safety sector that has been considered in this paper is the independent operability feature between the ALAGAS2 cores. The proposed ALGAS2 system has four ALGAS2 cores. These four ALGAS2 cores sharing the same digital circuits architecture. Each one of the proposed ALGAS2 core is consists of a localized FLS unit and Independent Inclination Control Unit (IICU) as shown in Figure 4. Also, each one of these ALGAS2 cores is working independently. Subsequently, the decision-making mechanism output from each ALGAS2 cores is only function of the high-speed differential input signals from the NI unit and the received signals from the Sensor Information Unit (SIU). Hence, there is no centralized decision-making signals that can dominate the operation of these processing units.

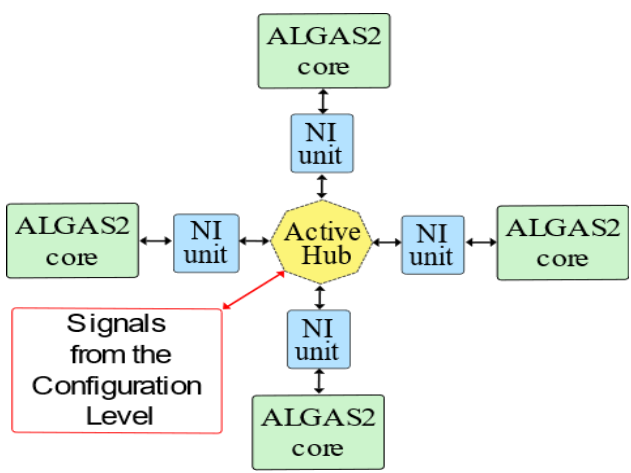

**Figure 3. A The block Diagram of the Proposed ALGAS2 system.**

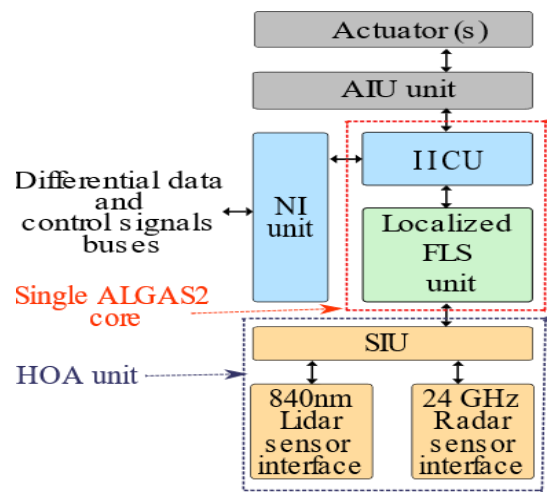

**Figure 4. A The block Diagram of the Proposed ALGAS2 core.**

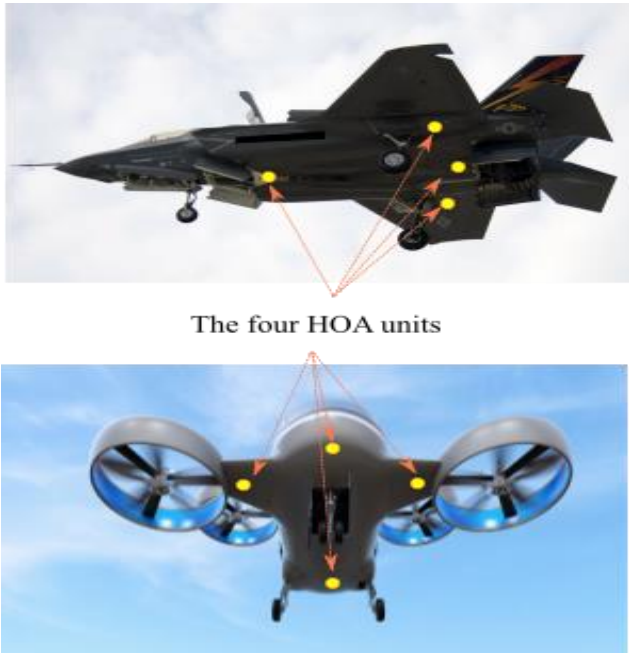

**Figure 5. A hypothetical integration of quad HOA modules on military and urban V\STOL drones.**

The main input signal source to each ALGAS2 core is from the HOA unit. The HOA unit is responsible for two main functions. The first function is related to the physical interconnections with the 840 nm lidar and the 24 GHz short range radar sensory modules. This task represents the smart connection level in the CPS-5C architecture. The second function is handled by the SIU. The SIU is responsible for performing the necessarily tasks at the data-to-information conversion level. This independent operations feature between the ALAGAS2 cores is enhancing the safety of a V/STOL landing operation since in case of a sudden failure occurred, the system could still handle this critical operation successfully. Also, the collaborative readings from the quad HOA units could be used to increase the accuracy of the measured distance between the drone and the landing spot.

The 3rd safety sector is referring to the usage of muti-spectral sensors, the radar and the lidar sensors. The large wavelength gap between these two sensors is adding more resilience and anti-jamming self-immunity during the landing operation, whether it is intentionally added or not. The main reason of selecting the combination between radar and lidar sensors is due to their high performance under almost any climate conditions while being covered by a protective radomes. The combination of all these proposed safety V/STOL landing techniques gives the proposed ALGAS2 system ultimate agility toward achieving a full-autonomous V/STOL drone network regardless to neither the type of landing spots nor the inclination angle between the landing surface and the drone's axes of motion in comparing with the conventional Inertial Measurement Units (IMU) sensors as illustrated in [7]. Moreover, these proposed ideas will have a significant impact toward introducing new V/STOL safety definitions for the next generation of UAM such as low-experience pilot UAM drones, no-experience pilot UAM drones, and even pilot-less UAM drones. We are completely convinced that this will be the anchor point toward expanding this technology globally. The selection of having four segregated ALGAS2+ HOA pairs is only to make sure that the overall ALGAS2 system is suitable to fit well in most of the futuristic V/STOL landing drones, whether they are space vehicles as illustrated in Figure 1, or urban/ military drones as depicted in Figure5.

**Table 1. Design Summary Report for The Proposed ALGAS2 Processing system Architecture Using Quartus Prime Tool.**

| Family name | Cyclone V |
|---|---|
| Device | 5CGXFC9D6F27C7 |
| Total logic elements in ALM | 3,488 ALMs |
| No. of ALGAS2 cores in the ALGAS2 system | 4 cores |
| External memory usage | None |
| Total DSP blocks | None |
| Maximum frequency | 252.40 MHz @1.1v &85C fast model 279.25 MHz @1.1v &85C fast model |
| No. of parallel operations per clock cycle | 76 |
| Maximum Giga Operations Per second (GOPS) | 21.22 GOPS |
| Junction temperature range | 0 to 85 ºC |
| Selected cooling solution | 23 mm heatsink with 200 LFpM airflow |
| Core dynamic thermal power dissipation | 146.4 mW |

Table 2. Proposed ALGAS2 architecture synthesis comparison with Ref. [7, 27,29-30].

|  | The Proposed ALGAS2 architecture in this paper | The Proposed ALGAS architecture in Ref. [7] | The Proposed architecture in Ref. [27] | The Proposed architecture in Ref. [29] | The Proposed architecture in Ref. [30] |
|---|---|---|---|---|---|
| FLS engines in the systems | 4 (Systolic-basic cores + independently operated) | 5 (systolic-based core) | 3 (systolic-based core) | 1 (systolic-based core) | 2 |
| Total dynamic power per system | 146.4 mW | 178.12 mW | 58.56 mW | 29.09 mW | 100.64 mW* |
| Memory usage | None | None | None | None | None |
| Max. freq. | 279.25 MHz | 266.03 MHz | 270.86 MHz | 181.55 MHz | 75.1 MHz |
| FLS crisp inputs resolution | Variable (11-bits, 10-bits) | Variable (11-bits, 10-bits) | Variable (11-bits, 10-bits) | Variable (11-bits, 10-bits) | 12 bits |
| No. of logic elements | 3,488 ALMs | 4,304 ALMs | 2,578 ALMs | 1641 ALMs | 672 LE |
| Maximum Giga Operations Per second (GOPS) | 21.22 GOPS | 25.273 GOPS | 14.36 GOPS | 3.085 GOPS | NA |

Furthermore, there is no specific standards to limits neither the engines types to be used for these V/STOL drones, nor the mechanical mechanism to be used. This was one of the drawbacks of the proposed V/STOL ALGAS system in *[7]*, since it cannot be used for some drones which have nozzle(s) at their center of gravity or equilibrium centers. Hence, it was obvious to propose a new ALGAS2 system for V/STOL drones that depends on four ALGAS2 cores instead of depending on an ALGAS system with four ALGAS cores in addition to one reference ALGAS unit.

## RESULTS

The most essential concern in designing the proposed ALGAS system is to optimize the dynamic power consumption rate since it could hinder the overall flight-time abilities for the drones that will adopt this system. One the other hand, it is so important not to negatively affect the overall computational performance so we can sustain high agility and accuracy levels during the V/STOL operations. To obtain these requirements, the proposed ALGAS2 system went through many iteration and modification stages at design stage of the localized FLS unit engines using MATLAB and at the digital circuits design for each block in the ALGAS2 system. After achieving the desired satisfactory tuning level and acceptable parameters of the FLS engines from the MATLAB, then we started to describe it using VHDL. The digital circuits architecture of the FLS engine is depending on a systolic processing technique to accelerate the computational performance. Also, the proposed FLS architecture has different data and control bus widths such as 7 bits, 8bits, 9 bits, 10 bits, 11 bits, 14 bits, and 15 bits. Subsequently, we removed any unnecessary bits that will not carry out any useful information between specific sub-blocks at this digital circuits design optimization stage.

The proposed FLS engine went through deep analyses between the results from MATLAB and the QuestaSim simulation tool. The results showed a perfect correlated data output with a maximum reading difference error less than 3%. This task has been accomplished by using 12 different samples from the MATLAB and comparing them with the corresponding results from QuestaSim simulation tool. After the FLS passed the optimization tests successfully, we integrated the IICU block to the each FLS engine to form an ALGAS2 core. The proposed ALGAS2 system is consists of four segregated ALGAS2 cores in addition to the simple active hub and four NI units.

The final analysis of the entire ALGAS2 system showed that it only consumed about 3.07% of the overall available logic block resources available in the Cyclone V 5CGXFC9D6F27C7 FPGA chip as shown in Table 1. Moreover, we can observe that the maximum operating frequency could reach out nearly 279.25 MHz. Subsequently the proposed ALGAS2 system could sustain a maximum computation performance around 21.22 GOPS under only 146.4 mW dynamic power consumption. This can be considered as a very satisfied computational performance rate for having real-time processing abilities in such critical V/STOL operations.

To secure the agility of the proposed ALGAS2 system against any cyber-attacks, the ALGAS2 system is not exchanging any data with any external memories. Also, the proposed ALGAS2 system is not depending on any generic Intellectual Property (IP) in order to make it much either to be implemented on different FPGA chips from different vendors, or as an ASIC chip without been hindered by any restriction or limitation of acquiring the rights-of-usage of these Ips form from third parties.

The proposed ALGAS2 system is mainly depends on the FLS engine to gain its agility and sustainable performance. Thus, we compared the achieved performance of it with similar FLS-based systems such as in [7, 27, 29, 30] as depicted in Table2. We could observe that the proposed ALGAS2 system is consuming more dynamic power than Ref. [27, 29-30] and less than the value achieved in Ref. [7]. This is directly related to the number of the FLS engines that have been used in each system. It is obvious also that the corresponding maximum operating frequency and the amount of the logic cell utilization in each one of the proposed systems is proportional to the number of FLS engines that have been used as well.

It is noticeable that the proposed ALGAS2 system in this paper has a better power consumption rate in comparing with the ALGAS system in Ref. [7] by approximately 17.80%. Also, there is an improvement in the overall logic cell consumption between them by about 18.96%. in spite of the degradation in the overall maximum computation performance from 25.273 GOPS in Ref. [7] to 21.22 GOPS, in the proposed ALGAS2 system, the overall attributes that have been possessed by the proposed ALGAS2 system could be perfectly matching the needs of any futuristic AI-based Decision Support Systems DSS as a subsystem in a more sophisticated system that can maintain more features for wide range of UAM Vehicle projects.

## Conclusion

The proposed ALGAS2 system is mainly depends on four segregated FLS processing engines. A systolic digital circuits architecture using VHDL has been used to design the FLS processing engine to boost its computational performance. The achieved 21.22 GOPS of the overall ALGAS2 computational performance is extremely beneficial for accomplishing a precise autonomous landing task for wide ranges of UAM vehicles. Many optimization and modifications processes will be introduced to the FLS engine itself to enhance its computational performance. Also, the digital circuits architecture of the current active hub unit and the NI unit need to be modified to accept a transfer data rate around 10 Gbps.